\documentclass[aps, prl,twocolumn,superscriptaddress]{revtex4-2}

\usepackage{siunitx}
\usepackage{graphicx}
\usepackage{amsmath, amssymb}
\graphicspath{{Figures/}}

\DeclareMathOperator{\sign}{sign}

\newcommand{\ud}{\mathrm{d}} 
\newcommand{\vect}[1]{\mathbf{#1}} 
\newcommand{\onedvect}[1]{\tilde{#1}} 

\begin{document}

\title{System size and boundaries determine the patterning dynamics of attracting active particles}

\author{Jan Rombouts}
\email{jan.rombouts@ulb.be}
\thanks{Equal contribution}
\affiliation{Cell Biology and Biophysics Unit, EMBL Heidelberg}
\affiliation{Developmental Biology Unit, EMBL Heidelberg}
\affiliation{Unit of Theoretical Chronobiology, Université Libre de Bruxelles}
\author{Michael L Zhao}
\thanks{Equal contribution}
\affiliation{Developmental Biology Unit, EMBL Heidelberg}
\author{Alexander Aulehla}
\affiliation{Developmental Biology Unit, EMBL Heidelberg}
\author{Anna Erzberger}
\email{anna.erzberger@embl.de}
\affiliation{Cell Biology and Biophysics Unit, EMBL Heidelberg}
\affiliation{Department of Physics and Astronomy, Heidelberg University}

\begin{abstract}
Pattern formation often occurs in confined systems, yet how boundaries shape patterning dynamics is unclear. We develop techniques to analyze confinement effects in nonlocal advection-diffusion equations, which generically capture the collective dynamics of active self-attracting particles. We identify a sequence of size-controlled transitions that generate characteristic slow modes, leading to exponential increase of patterning timescales. Experimental measurements of multicellular dynamics confirm our predictions.

\end{abstract}

\maketitle

\paragraph*{Introduction.}

Attractive interactions between the constituents of active systems lead to rich collective phenomena, ranging from assemblies of active colloids \cite{theurkauff_dynamic_2012, palacci_living_2013} to the formation of animal groups \cite{vicsek_collective_2012, ioannou_multiscale_2023}. In multicellular systems, pattern formation through cell aggregation appears in many different situations, from the formation of slugs in \textit{Dictyostelium} \cite{weijer_dictyostelium_2004a} to the recruitment of immune cells to a site of infection \cite{strickland_selfextinguishing_2024}. Self-organized aggregation is central to development \cite{ros-rocher_chemical_2023,abitua_axis_2024}, synthetic patterning \cite{mcnamara_recording_2024, fiuza_morphogenetic_2024, toda_programming_2018}, and possibly the evolution of multicellularity \citep{brunet_origin_2017}. At larger spatial scales, interactions between individual organisms lead to the collective structures such as fish schools, bird flocks or human crowds.
Although these processes often take place in confined environments, how boundary interactions and geometrical constraints affect aggregation-induced pattern formation is not well understood.

Nonlocal advection-diffusion models offer a powerful framework for describing the dynamics of self-attracting constituents without requiring detailed knowledge of interaction mechanisms. By coarse-graining molecular, chemical, or mechanical interactions into an effective kernel, they can capture a broad class of physical and biological systems \cite{painter_biological_2024}. While such models have been applied from subcellular to ecological scales, most studies focus on infinite or periodic domains (but see Refs. \cite{hillen_nonlocal_2020,yu_directed_2025}), and direct comparisons with experimental data remain rare.

In this Letter, we investigate how confinement alters pattern formation in a minimal nonlocal model of attractive active particles. We show that increasing system size triggers a sequence of bifurcations that generate slow dynamical modes, leading to an exponential increase of the patterning timescale. These theoretical predictions are confirmed by experimental data of primary reaggregated murine cells which form patterns on size-controlled substrates. Our results reveal how finite system size and boundary interactions control the timescale of pattern formation driven by long-range attraction.

\paragraph*{Non-local advection-diffusion dynamics.}

\begin{figure}[!t]
\includegraphics[width=\columnwidth]{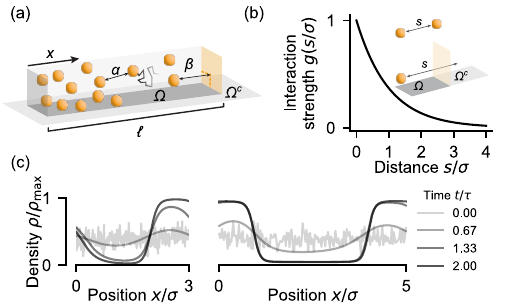}\caption{\textbf{Pattern formation of self-attracting particles in finite-sized systems.} In a one-dimensional system of length $\ell$, active particles (orange) undergo random and directed motion due to interparticle coupling $\alpha$ and particle-boundary coupling $\beta$. b) We describe interparticle and particle-boundary interactions with nonlocal exponential kernels with characteristic decay length $\sigma$. c) Simulations of Eq.~\eqref{eq:scaled} illustrate the patterning dynamics for two different domain lengths. ($\alpha=18, \beta=7.2, \rho_0=0.4$)}
    \label{fig:introduction}
\end{figure}

We investigate the influence of finite system size and boundaries on the collective dynamics of self-attracting particles whose density field $\rho(\vect{r}, t)$ evolves according to the mass-conserving advection-diffusion equation
\begin{equation}
    \partial_t \rho = \nabla\cdot\left(D\nabla\rho - \vect{v}(\vect{r}) \rho\right).
\end{equation}
The particles undergo random motion set by the diffusion coefficient $D$, and advective motion with velocity $\vect{v}(\vect{r})$ (Fig.~\ref{fig:introduction}a). We focus on flows driven by forces arising due to interactions between particles and with the boundaries, such that in the overdamped limit, the velocity field satisfies
\begin{equation}
   \gamma(\vect r)  \vect{v}(\vect{r}) =  \vect{f}_{\rm{p}}(\vect{r}) + \vect{f}_{\rm{b}}(\vect{r}), \label{eq:v_twoparts}
\end{equation}
in which we introduce $\vect{f}_{\rm{p}}(\vect{r})$ and $\vect{f}_{\rm{b}}(\vect{r})$ as the forces acting at $\vect{r}$ due to particle-particle and particle-boundary interactions respectively. The dissipative coefficient $\gamma(\vect r)$ is taken to depend on the local density, $\gamma(\vect r) = \gamma (\rho)$, such that higher densities impede particle movement. Here, we use $\gamma(\rho) = \gamma_0/(1-\rho/\rho_{\rm{max}})$, an expression derived from microscopic random walks with occupation-dependent jumping probabilities \citep{painter_volumefilling_2002}. We represent generic interaction forces using non-local kernels of the form
\begin{equation}
    \vect{f}_{\rm{p}}(\vect{r}) =\int_\Omega \vect{G}_\textup{p}(\vect{r'}-\vect{r}) \rho(\vect{r}') \ud\vect{r'}, \quad 
     \vect{f}_{\rm{b}}(\vect{r}) = \int_{\Omega^c} \vect{G}_{\textup{b}}(\vect{r}'-\vect{r})\ud \vect{r'}, \label{eq:forces}
\end{equation}
in which $\Omega$ is the domain containing the particles and $\Omega^{\rm{c}}$ is the space outside (Fig.~\ref{fig:introduction}a,b). 
More generally, a boundary interaction can be expressed as $\int \vect{G}_{\textup{b}}(\vect{r}'-\vect{r})\phi(\vect r' )\ud \vect{r'}$, where $\phi$ denotes a phase field defining the external material. Eq.~\eqref{eq:forces} is obtained for $\phi = 1$ outside the domain and $\phi=0$ inside. 
The interaction kernels are parameterized as
 \begin{equation}
     \vect G_i(\vect{r}'-\vect{r}) = \frac{\chi_i}{\sigma_i} g_i\left(\frac{|\vect{r}'-\vect{r}|}{\sigma_i}\right) \frac{\vect{r}'-\vect{r}}{|\vect{r}'-\vect{r}|}
     \label{eq:interactionkernel}
 \end{equation}
such that $\sigma_i$ and $\chi_i$ determine the length scale and amplitude of the interaction, and the normalized function $g_i$ describes how the magnitude of the force depends on distance, with $i \in \{\rm{p};\rm{b}\}$ (Fig.~\ref{fig:introduction}b). Different microscopic mechanisms can be shown to map to such interaction kernels. For example, Keller-Segel-type chemotactic attraction between motile cells \citep{keller_initiation_1970, painter_mathematical_2019} leads to an exponentially decaying interaction kernel with length scale $\sigma_\textup{p} = \sqrt{D_c / b}$, determined by the diffusion constant $D_c$ and the degradation rate $b$ of the chemoattractant \citep{lee_nonlocal_2001}. Similarly, for contractile fluids \citep{bois_pattern_2011, palmquist_reciprocal_2022}, the equivalent nonlocal description contains an exponential kernel with decay length corresponding to the hydrodynamic lengthscale $\sqrt{\eta/\gamma}$, with $\eta$ the viscosity and $\gamma$ a friction coefficient \cite{supplementalmaterials}. 

Considering interaction kernels with equal shape and range for the inter-particle and boundary-particle interactions in a one-dimensional system, we obtain the equation 
 \begin{equation}
 \begin{aligned}
     \partial_t \rho = \partial_{x}\biggl( \partial_x\rho - \rho (1-\rho) \biggl[ &\alpha \int_0^\ell \onedvect{g}(x'-x) \rho(x') \ud x' \\
     + &\beta \int_{\mathbb{R}\backslash[0,\ell]}\onedvect{g} (x'-x) \ud x'\biggl]\biggr),\label{eq:scaled}
     \end{aligned}
 \end{equation}
in which space, time, and the density have been rescaled by $\sigma, \tau = \sigma^2/D$ and $\rho_\text{max}$ respectively \cite{supplementalmaterials}, the system size is given relative to the interaction range $\ell=L/\sigma$, and $\onedvect{g}(x'-x) = \sign(x'-x)g(|x'-x|)$, and the external material is assumed constant in time.

 The dynamics are controlled by the dimensionless particle-particle interaction parameter $\alpha = \chi_\textup{p} \rho_\text{max} \sigma/(\gamma_0D)$ and the particle-boundary interaction parameter $\beta=\chi_b\sigma/(\gamma_0D)$, as well as the total number of particles, which determines  $\rho_0 = \big(\int_0^\ell\rho(x)\ud x \big)/\ell$.

Fig.~\ref{fig:introduction}c shows simulations of Eq.~\eqref{eq:scaled} with no-flux boundary conditions, illustrating how spontaneous patterns form in this system. 

\paragraph*{Conditions for patterning}

On periodic or infinite domains, patterns form spontaneously from a homogeneous particle distribution if \cite{supplementalmaterials,jewell_patterning_2023}
\begin{equation}
     \alpha \rho_0 (1-\rho_0)  > \frac{1}{2}. \label{eq:patterningthreshold_infinitedomain}
\end{equation}
Thus, since $\alpha$ represents the relative strength of attraction compared to diffusion, patterning is possible if attraction overcomes random motion and if the average density is neither too low nor too high. 

\begin{figure}
    \includegraphics{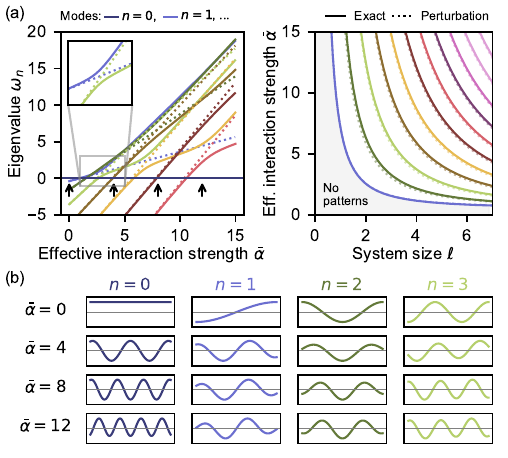}\caption{\textbf{Linear stability analysis of the nonlocal equation in finite-sized systems. } a) Approximations (dotted lines, Eq.~\ref{eq:omega_perturbation}) and exact values (solid lines, numerical solutions of Eq.~6 in \cite{supplementalmaterials}) of the eigenvalues $\omega_n$ show that eigenvalue curves (`loci') undergo turning accompanied by the cross-over of their approximations (inset). The instability lines (right) of each spatial mode along which $\omega_n = 0$ are well-approximated analytically.  b) The shape of the eigenmodes changes with the parameters, here shown for different values of $\bar\alpha$ (arrows in panel a).}
    \label{fig:linear}
\end{figure}

While all uniform profiles $\rho(x) = \rho_0$ are steady states on infinite domains, this is not the case on bounded domains, for which instability thresholds are analytically accessible only in specific limits. In particular, considering a \emph{neutral} boundary interaction with $\beta = \alpha\rho_0$ allows linearizing Eq.~\eqref{eq:scaled} around a uniform steady state $\rho(x) = \rho_0$, leading to the eigenvalue problem \cite{supplementalmaterials}
\begin{equation}
     (\partial_{xx} - \bar\alpha\partial_x\mathcal{A}) u = \omega u, \qquad (\partial_x u - \bar\alpha \mathcal{A} u )|_{x \in \{0;\ell\}} = 0 
     \label{eq:eigenu}
\end{equation}
in which 
\begin{equation}
     \bar\alpha = \alpha\rho_0(1-\rho_0)\quad\text{and}\quad(\mathcal{A} u)(x) = \int_0^\ell \onedvect g(x'-x) u(x') \ud x'.
\end{equation}
The solutions of this equation yield the eigenfunctions $u$ and associated growth rates $\omega$ (Fig.~\ref{fig:linear}). In contrast to infinite or periodic domains, these spatial modes are in general not sinusoidal: their shape depends on the domain length and the interaction parameters (Fig.~\ref{fig:linear}b).
For small $\bar\alpha$, Eq.~\eqref{eq:eigenu} can be written as a perturbation of the eigenvalue equation for the Laplacian. The first-order perturbative calculation approximates the $n$th eigenvalue as
\begin{equation}
     \omega_n \approx \omega_n^0 + \bar\alpha \omega_n^1, \label{eq:omega_perturbation}
\end{equation}
in which
\begin{equation}
\begin{aligned}
\omega_n^0 &= - \left(\frac{n\pi}{\ell}\right)^2, \quad
\omega_n^1 = \frac{\int_0^\ell (u^0_n)'(x) (\mathcal{A} u_n^0)(x) \ud x}{\int_0^\ell u_n^0(x)^2 \ud x} ,\\
 u_n^0 &= \cos\left(\frac{n\pi}{\ell}x\right).
    \end{aligned}
\end{equation}
To evaluate the accuracy of this approximation, we seek a class of interaction kernels for which exact solutions can be obtained. We find that for kernels of the form $g(s)\in~\mathrm{span}\{s^m e^{-s}\ |\  m~\in~\mathbb{Z}_{\geq0}\}$, a solution $u$ of Eq.~\eqref{eq:eigenu} is also a solution of a linear, constant-coefficient ordinary differential equation \cite{supplementalmaterials}. For $m=0$, i.e. an exponentially decaying kernel, an eigenfunction $u$ satisfies
\begin{equation}
 \frac{\ud^4u}{\ud x^4} + \left(2\bar\alpha - 1 - \omega\right) \frac{\ud^2 u}{\ud x^2}+\omega u = 0 .\label{eq:ode}
\end{equation}
Equations \eqref{eq:ode} and \eqref{eq:eigenu} yield algebraic equations for the eigenvalues and expressions for the spatial modes. We find that the curves describing the eigenvalues cross and turn as $\bar{\alpha}$ increases, with the eigenmodes changing shape accordingly, and the approximations following straight lines (Fig.~\ref{fig:linear}a-b) --- a phenomenon reminiscent of `eigenvalue veering', usually studied in structural dynamics \cite[e.g.][and references therein]{manconi_veering_2017}. 
Moreover, identifying for which parameters the growth rate of the $n$th spatial mode $\omega_n$ becomes positive reveals the instability curves in the $(\ell,\bar\alpha)$-plane as solutions of \cite{supplementalmaterials}
\begin{equation}
     e^{\ell z} =  (-1)^n \frac{z-1}{z+1} \quad\text{with}\ z = \sqrt{1-2\bar\alpha},\label{eq:instabilitycurves}
\end{equation}
in which $n$ indicates the parity of the spatial mode (Fig.~\ref{fig:linear}a, right). 

Comparing the exact solution to the approximation Eq.~\eqref{eq:omega_perturbation}, we find that the latter does not capture the intricate behavior of the eigenvalues for higher values of $\bar\alpha$, but agrees well in the regime where $\omega_n<0$, and provides an excellent approximation of the patterning thresholds (Fig.~\ref{fig:linear}a).

\paragraph*{Long-time solutions are polarized for a wide range of boundary interactions.}
To study the long-time dynamics, and to analyze a broader range of boundary interactions, we perform a numerical bifurcation analysis of the 
steady states of Eq.~\eqref{eq:scaled} as a function of the inter-particle interaction $\alpha$, the domain length $\ell$ and the particle-boundary interaction $\beta$. To characterize the patterns, we introduce an order parameter $\mu(\rho) = A(\rho) ||\rho-\rho_0||_2$, given by the $L_2$-norm of the profile, multiplied by the prefactor $A(\rho) = \int_0^\ell \left( \frac{1}{2} (x-1/2) + (x-1/2)^2\right)(\rho(x) - \rho_0)\ud x $, which distinguishes a solution $\rho(x)$ from its horizontally and vertically mirrored versions. 

\begin{figure}
\includegraphics{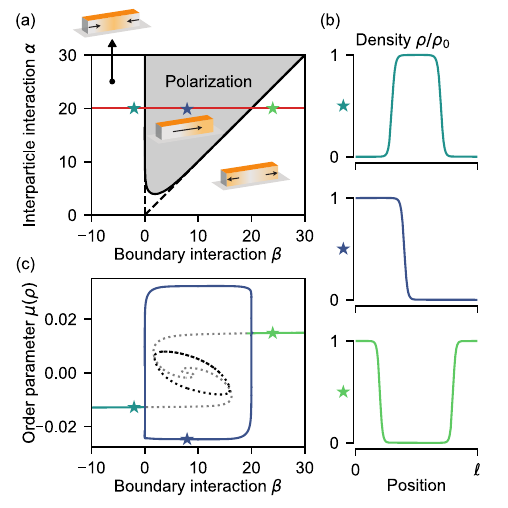}
    \caption{\textbf{Boundary interactions determine the final density patterns.} a) Increasing interparticle attraction $\alpha$ enlarges the parameter regime yielding polar final patterns (gray), whereas repulsive boundaries ($\beta < 0$), and strongly attracting boundaries ($\beta > \alpha$) produce symmetric patterns with peaks at the center or boundaries. (Solid black lines: numerically obtained pitchfork bifurcation lines. Dashed lines: $\beta=0$ and $\beta=\alpha$. Parameters: $\rho_0=0.4,\ell=5$.) b) Steady-state density profiles for the star-marked points in panel (a) show the symmetric and polar final patterns. c) 
    The bifurcation diagram along the red line in (a) ($\alpha=20$) reveals multiple unstable multi-peak states (dashed lines) inside the polar regime. )
    \label{fig:boundary}
    }
\end{figure}

We find that repulsive boundary interactions ($\beta<0$) always result in a single density peak at the center of the domain, while strongly attracting boundaries (i.e. $\beta >\alpha$) deplete the particles from the center, producing symmetric profiles with high densities at the boundaries. Weakly attractive boundaries ($0<\beta<\alpha$) generate polarized steady states ---asymmmetric profiles with a single high-density region at either boundary (Fig.~\ref{fig:boundary}).  

\begin{figure}
\includegraphics{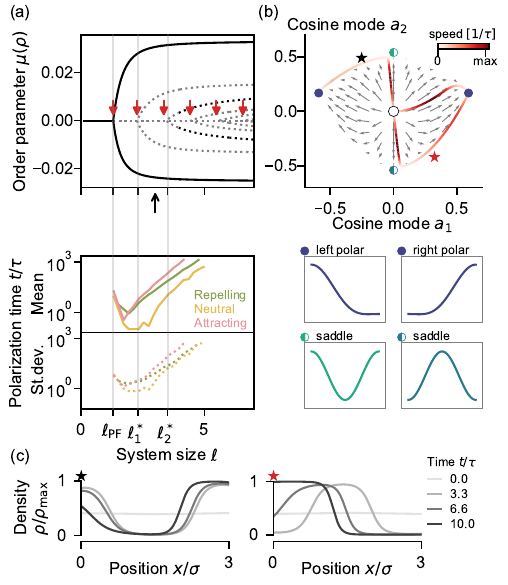}
\caption{\textbf{Increasing the system size triggers the appearance of saddle points that delay polarization by inducing slow transients.} a) Patterns evolve towards stable polarized states for domain sizes above the critical value $\ell_{\rm{PF}}$. These polar solutions remain the only stable steady states for larger values of $\ell$, but multiple saddle branches appear via a sequence of pitchfork bifurcations, giving rise to rich transients which support multi-peaked patterns. Transition points are well approximated by the analytical instability thresholds determined by the linear stability analysis (red arrows). Parameters: $\alpha=20, \rho_0=0.4$, $\beta = \alpha \rho_0 = 8$.  Statistics of 100 simulations show that both the average polarization time and its standard deviation depend exponentially on the domain size (lower panel). Neutral boundary interactions ($\beta=\rho_0\alpha$) promote faster polarization than repulsive (here: $\beta= 1/2\rho_0 \alpha$) or attractive ($\beta = 3/2\rho_0\alpha$) boundaries (Fig.~S4, \cite{supplementalmaterials}).
b) A phase portrait of the first two cosine modes $a_1$ and $a_2$ illustrates how the saddle points guide the transient dynamics. Depending on the initial conditions, solutions can be attracted to one of the symmetric transient patterns where they spend a long time before slowly polarizing. (Parameters:  $\alpha=20, \ell=3, \rho_0=0.4$, black arrow in panel a).  c) Snapshots of the density profiles from two trajectories in panel b (stars) show polarization via peak depletion and peak motion. 
}
    \label{fig:saddles}
\end{figure}

These are the only \emph{stable} steady states, but other solutions are present within the polarizing regime, associated with solutions that have even or odd symmetry around $\ell/2$ (Fig.~\ref{fig:boundary}c, dashed lines). We find that the presence of these unstable states depends on the domain size: as $\ell$ increases, the system undergoes a series of pitchfork bifurcations at each of which additional higher order spatial modes appear (Fig.~\ref{fig:saddles}a, Fig.~S1 and see Fig.~S3 for a similar dependence on $\alpha$).
These states correspond to saddle points, which --- albeit unstable ---can strongly influence the patterning \emph{dynamics}. In particular, these points generate slow dynamics on their unstable manifold (Fig.~S2, S3). Depending on the initial conditions, density profiles may quickly approach one of these saddles, before slowly resolving into the polarized state, possibly passing by other saddle points. The parameter values at which these solutions appear correspond to the instability thresholds for successive spatial modes, and are thus well approximated by solving Eq.~\eqref{eq:omega_perturbation} for $\omega_n=0$ and given $\bar\alpha$ (red arrows in Fig.~\ref{fig:saddles}a).

For $\alpha=20$ and $\ell\in(\ell^*_1,\ell^*_2)$, four patterned steady states exist, and we can project the density profile onto the first two cosine modes to parameterize a phase plane that shows the symmetric and antisymmetric saddle points and their associated dynamics (Fig.~\ref{fig:saddles}b, c). For trajectories that first approach the middle-peak saddle, the density profile develops a central peak, which subsequently slowly moves to the side (Fig.~\ref{fig:saddles}c, right), whereas passing by the two-peak saddle leads to one peak slowly losing material to the other one, a phenomenon similar to Ostwald ripening in phase-separating systems (Fig.~\ref{fig:saddles}c, left) \cite{voorhees_ostwald_1992}.  Coarsening from multi-peak transients to single-peak final states is a generic feature observed in many mass-conserving pattern-forming or phase-separating systems \citep{weyer_coarsening_2023,potts_distinguishing_2024,argentina_coarsening_2005}. 

Increasing the system size or the strength of inter-particle interactions triggers the appearance of more branches with saddle points (Fig.~\ref{fig:saddles}a, Fig.~S1, S3). To investigate how the number of saddle-branches influences the duration and variability of trajectories, we performed numerical simulations with random initial conditions for different system sizes and measured the time it takes the system to polarize. We find that both the average and standard deviation of the polarization time $\tau_{\rm{pol}}$ (defined as the timepoint from which the first cosine mode is the dominant contribution to the density profile) increases exponentially with system size, with neutral boundaries promoting more rapid polarization than repelling or attractive ones (Fig.~\ref{fig:saddles}a lower panel, Fig.~S4). Thus, confinement to systems shorter than the critical length $\ell^*_1$ at which the saddle branches appear guarantees the fast and robust polarization, whereas the dependence on initial conditions becomes more pronounced in larger systems.

\begin{figure*}
\includegraphics{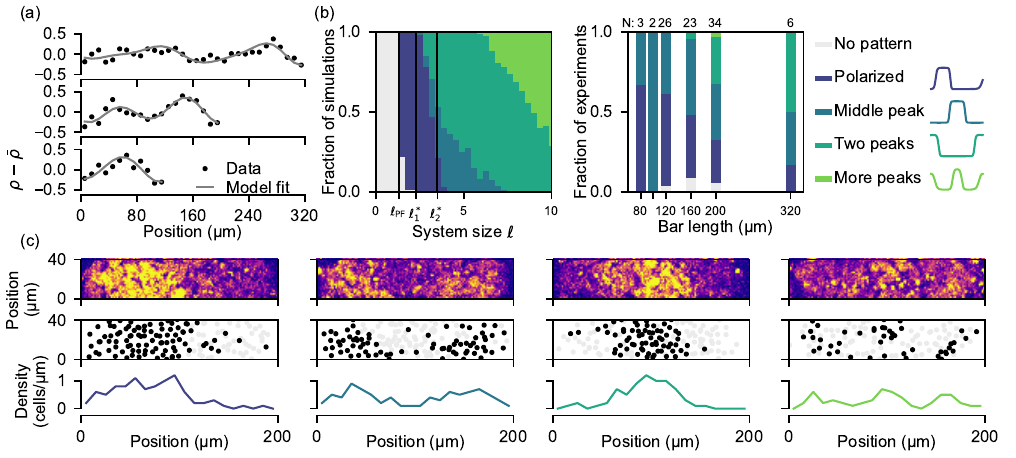}
    \caption{\textbf{Experimental patterns of self-attracting cells on size-controlled substrates show the predicted distribution of transients. } a) Embryonic mouse tailbud primary cells form aggregation patterns on size-controlled microprinted bars \cite{preprint_michael}. We fit the effective interaction strength $\bar\alpha$ (and $\alpha$) to the early density evolution with the linearized Eq.~\ref{eq:scaled} (\cite{supplementalmaterials}, Fig.~S6). b) The distribution of transient patterns from 200 simulations (left) and experimental measurements with sample size $N$ (right) show characteristic shifts as the system size is increased ($\alpha=20,\rho_0=0.4, \beta=\alpha\rho_0$, \cite{supplementalmaterials}, Fig.~S7). 
    c) Experimental examples show distinct pattern types. The density profiles of high-Brachyury cells were obtained by quantifying their positions (black dots in middle panels) from Brachyury fluorescence micrographs (top panels: maximum intensity projections, see Fig.~S5).}
   
    \label{fig:experimental}
\end{figure*}

\paragraph*{Experimentally observed patterns show signatures of saddle-induced transients.}
In complex systems such as living matter, developing patterns can be arrested, read out, or modified by other processes, depending on their relative timing \cite{ebisuya_what_2018, garcia-ojalvo_time_2023}. \emph{Transient dynamics} can thus play an important role, also in establishing final structures \cite{koch_ghost_2024}. To investigate whether the predicted dynamics can be observed experimentally, we performed experiments using primary cells from mouse embryos, which form aggregation patterns on rectangular microprinted substrates (Fig.~\ref{fig:experimental}, \cite{tsiairis_selforganization_2016, ho_nonreciprocal_2024a, preprint_michael}). We quantified one-dimensional density profiles of cells expressing Brachyury, a mesodermal marker involved in different symmetry-breaking events in early development \citep{sebe-pedros_early_2013, schwaiger_ancestral_2022}, from the imaging data (Fig. S5, \cite{supplementalmaterials}). By fitting the linearized model to the initial timepoints of these experiments, we estimated an interaction strength of roughly $\alpha\approx 20$  (Fig.~\ref{fig:experimental}a, Fig.~S6). Next, we examined whether using this parameter value, the theoretical distribution of transient patterns corresponds to experimentally observed intermediate patterns. 
We simulated Eq.~\ref{eq:scaled} for varying system sizes $\ell$ from 200 random initial conditions per value, and classified the density profiles at the timepoint $2.4 \tau \ll \tau_{\rm{pol}}$. As expected, we observe a distinct increase in the frequency of symmetric and multi-peaked profiles for larger systems (Fig.~\ref{fig:experimental}b). We performed the same pattern classification on the experimental data at \SI{16}{\hour} post plating (comparable to $2.4 \tau$) and indeed observe the predicted shifts in the frequency of polarized, symmetric, and multi-peaked profiles (Fig.~\ref{fig:experimental}b,c). In particular, we find that confining the system to domains smaller than \SI{160}{\micro\meter} suppresses the formation of multiple peaks, highlighting how system size can control and guide pattern formation. 

\paragraph*{Conclusions.}

Confinement influences collective behavior in biological, physical, and even social systems \cite{araujo_steering_2023}. We developed analytical and numerical tools to investigate pattern formation in nonlocal advection-diffusion equations on bounded domains. 
For exponentially decaying interactions, we showed that increasing the system size triggers a sequence of transitions that generate slow dynamical modes and an exponentially increasing polarization time. Our analysis provides mechanistic insight into active coarsening dynamics in a low-dimensional phase space, controlled by saddle points and their manifolds. It will be interesting to explore how these phase-space structures shape dynamics in nonequilibrium systems with active turnover \citep{brauns_wavelength_2021,potts_distinguishing_2024,zwicker_suppression_2015} or growth-induced changes in size \citep{miguez_effect_2006,konow_turing_2019}.

While we analyzed a specific kernel, our methods extend to systems with other nonlocal interactions, potentially including neural fields \cite{coombes_neural_2014}, dispersal in ecology \cite{pueyo_dispersal_2008} or kernel-based reaction-diffusion models \cite{ei_effective_2021}, as well as systems with kernels inferred directly from experimental data \cite{frishman_learning_2020, hiscock_mathematically_2015}.

\paragraph*{Acknowledgments}
We are thankful to Hanspeter Herzel, Geneviève Dupont and all the members of the Erzberger group for manuscript feedback. We thank Alf Gerisch and Daniele Avitabile for advice on numerical aspects. Special thanks go to Jordi Garcia-Ojalvo, for discussions, feedback and advice throughout the development of this project. 
JR was supported by an EMBL
Interdisciplinary Postdoctoral Fellowship (EIPOD4) program under Marie Sklodowska-Curie Actions Cofund (grant agreement no. 847543) and a postdoctoral fellowship from FNRS (Chargé de recherches). MLZ was supported by a Bridging Exellence Fellowship from the EMBL|Stanford Life Science Alliance. We acknowledge the use of the EMBL High Performance Computing resources \cite{emblhpc}. We thank the Advanced Light Microscopy Facility (ALMF) at the European Molecular Biology Laboratory (EMBL) and Zeiss for support.

\nocite{imuta_tknockin_2013}
\nocite{abe_rosa26_2011}
\nocite{weigert_stardist3d_2020}
\nocite{berg_ilastik_2019}
\nocite{gerisch_approximation_2010}
\nocite{trefethen_spectral_2000}
\nocite{kuznetsov_elements_2004}
\nocite{seydel_practical_2010}
\nocite{painter_mathematical_2019}
\nocite{bois_pattern_2011}
\nocite{palmquist_reciprocal_2022}
\nocite{jewell_patterning_2023}

\bibliography{patterningboundaries_bibliography}

\end{document}